\documentclass[12pt,fleqn,twoside]{article}
\usepackage{amsmath}
\usepackage{amsfonts}
\usepackage{espcrc1}
\usepackage{graphicx}

\input{tcilatex}

\begin{document}

\title{Hamiltonian treatment of time dispersive and dissipative media within the linear response theory}%

\author{Alexander Figotin\address{Department of Mathematics, University of California at Irvine, Irvine, CA 92697}
and
Jeffrey Schenker\footnote{current address: School of Mathematics, Institute for Advanced Study, Princeton, NJ 08540 }\address{Institut f\"ur Theoretische Physik, ETH Z\"urich, CH-8093
Z\"urich, Switzerland}}%

\maketitle%

\begin{abstract}%
We develop a Hamiltonian theory for a time dispersive and dissipative (TDD)
inhomogeneous medium, as described by a linear response equation respecting
causality and power dissipation. The canonical Hamiltonian constructed here
exactly reproduces the original dissipative evolution after integrating out
auxiliary fields. In particular, for a dielectric medium we obtain a simple
formula for the Hamiltonian and closed form expressions for the energy
density and energy flux involving the auxiliary fields. The developed
approach also allows to treat a long standing problem of scattering from a
lossy non-spherical obstacle and, more generally, wave propagation in TDD
media.
\end{abstract}%

\section{Introduction}

There is an intrinsic ambiguity in identifying the field energy densities
for radiation in a time dispersive and dissipative (TDD) medium as described
by the linear response theory, e.g., in a dielectric medium described by the
classical linear Maxwell equations with complex valued frequency dependent
electric permittivity $\varepsilon (\omega )$ and magnetic permeability $\mu
(\omega )$. Consequently, there are problems with the interpretation of the
energy balance equation \cite[\S 77]{LandLifEM}, \cite[\S 1.5a]{FelMarc},
\cite[\S 6.8]{Jackson}, \cite{Loudon}. There were a number of efforts \cite%
{Loudon}, \cite{LaxNelson}, \cite{NelsonChen} to construct a consistent
macroscopic theory of dielectric media that accounts for dispersion and
dissipation, based on more fundamental microscopic theories. At first sight, it
seems that the introduction of a realistic material medium in an explicit form
similar to \cite{Loudon}, \cite{LaxNelson}, \cite{NelsonChen} is the only way
to model a TDD medium. In fact, that is not so and in this paper we describe a
consistent macroscopic approach \textit{within the linear response theory}.
Full proofs of the statements outlined here will appear in a forthcoming paper
\cite{FS2}.

A linear response TDD medium is an essentially open dissipative system, which
in principle can be obtained by \textit{i.}) eliminating some degrees of
freedom from a more involved microscopic theory and \textit{ii.}) making the
approximation of linear response. Stopping short of introducing a microscopic
theory we ask, \textit{ is there a conservative extended system which exactly
reproduces the given linear TDD system after reduction?} In \cite{FS} we showed that
indeed such an extension is \textit{i.}) possible and \textit{ii.}) essentially
uniquely determined, under general conditions of causality, power dissipation,
and minimality of the extension. Here we go further and construct a canonical
Hamiltonian for such a conservative extension based only on the given TDD
equations \textemdash \ without assumption on the underlying microstructure. In
particular, we construct such a Hamiltonian for a dielectric medium as defined
by complex $\varepsilon \left( \omega \right) $ and $\mu \left( \omega \right)
$. The construction given here is not restricted to a dielectric medium,
however, but holds for TDD systems with a certain mathematical structure
\textemdash \ eqs.\ (\ref{hpq4a}, \ref{hpq5}, \ref{PDC}) below \textemdash \
including, in particular, elastic and acoustic media, and it can be extended to
space dispersive dissipative systems. A somewhat related construction of the
evolution equations for linear absorptive dielectrics was given by Tip
\cite{Tip}. The range of validity of the proposed theory is the same as for the
linear response, though nonlinear generalizations are clearly possible.

Other important benefits of the approach developed here are: \textit{i.})
The constructed Hamiltonian is an integral of a local energy density, which
in the absence of TDD terms reduces to the local field energy. This permits
us to derive an expression for the energy transport for TDD media. \textit{%
ii.}) The present formulation allows to treat, in particular, a long
standing problem of scattering from a lossy nonspherical scatter -\ analyzed
by other methods with limited success \cite{Mishchenko} -\ by applying the
well developed scattering theory or conservative systems, see \cite{Newton,RS3} and references therein.
These applications will be discussed in detail in forthcoming work \cite{FS2},
\cite{FS3}.

\section{Construction of the Hamiltonian}
We consider a system described by two canonical vector coordinates $p,q\in H$%
, with $H$ a real Hilbert space. In the absence of TDD terms, the evolution
is supposed to be induced by a Hamiltonian $A\left( p,q\right) $ of the form
\begin{equation}
A\left( p,q\right) =\frac{1}{2}\left\langle K_{\mathrm{p}}p\,,\,K_{\mathrm{p}%
}p\right\rangle +\frac{1}{2}\left\langle K_{\mathrm{q}}q\,,\,K_{\mathrm{q}%
}q\right\rangle ,  \label{hpq1}
\end{equation}%
with closed linear operators $K_{\mathrm{p}}$, $K_{\mathrm{q}}$ from $H$
into auxiliary spaces $H_{\mathrm{p}}$, $H_{\mathrm{q}}$ respectively. To
manifest the conservation of energy, it is convenient to consider the
evolution of
\begin{equation}
f_{\mathrm{p}}:=K_{\mathrm{p}}p\in H_{\mathrm{p}},\quad f_{\mathrm{q}}:=K_{%
\mathrm{q}}q\in H_{\mathrm{q}},  \label{hpq3}
\end{equation}%
in place of $p,q$. In the absence of dissipation, these quantities evolve
according to
\begin{equation}
\partial _{t}%
\begin{pmatrix}
f_{\mathrm{p}} \\
f_{\mathrm{q}}%
\end{pmatrix}%
=%
\begin{pmatrix}
0 & -K \\
K^{\dagger } & 0%
\end{pmatrix}%
\begin{pmatrix}
f_{\mathrm{p}} \\
f_{\mathrm{q}}%
\end{pmatrix} , \quad \text{zero dissipation}
,  \label{hpq4}
\end{equation}%
with $K:=K_{\mathrm{p}}K_{\mathrm{q}}^{\dagger }$ a closed linear map from $%
H_{\mathrm{q}}$ to $H_{\mathrm{p}}$. Note that%
\begin{equation}
A(p,q)=\frac{1}{2}(\Vert f_{\mathrm{p}}\Vert ^{2}+\Vert f_{\mathrm{q}}\Vert
^{2})
\end{equation}
is conserved due to the antisymmetry of the generator in (\ref{hpq4}).

The electromagnetic field in a non-dispersive inhomogeneous medium may be
described in this framework, with $p=(4\mathrm{\pi })^{-1}\mathbf{A}$
(magnetic potential), $q=\mathbf{D}$ (electric displacement), $f_{\mathrm{p}%
}=(2\sqrt{\mathrm{\pi }})^{-1}\mathbf{H}$ (magnetic field), and $f_{\mathrm{q%
}}=(2\sqrt{\mathrm{\pi }})^{-1}\mathbf{E}$ (electric field). Identifying (%
\ref{hpq3}) with the material relations, we determine the action of the
operators $K_{\mathrm{w}}$:
\begin{equation}
K_{\mathrm{p}}\frac{\mathbf{A}}{4\mathrm{\pi }}(\vec{r})=(2\sqrt{\mathrm{\pi
}}){\mu }^{-1}(\vec{r})\cdot \left\{ \nabla \times \frac{\mathbf{A}}{4%
\mathrm{\pi }}(\vec{r})\right\} ,\qquad K_{\mathrm{q}}\mathbf{D}(\vec{r})=%
\frac{1}{2\sqrt{\mathrm{\pi }}}{\epsilon }^{-1}(\vec{r})\cdot \mathbf{D}(%
\vec{r}),  \label{ma1a}
\end{equation}%
where {$\mu $}, ${\epsilon }$ are the static permeability and dielectric
tensors, assumed real and symmetric. We take $(4\mathrm{\pi })^{-1}\mathbf{A}
$ and $\mathbf{D}$ in the space $H=H_{\mathrm{curl}}$ of divergence free
vector fields ---\ which amounts to a choice of gauge and an assumption of
no free charges. To complete the picture we define $H_{\mathrm{p}}$, $H_{%
\mathrm{q}}$ to be weighted $L^{2}$ spaces with scalar products
\begin{equation}
\left\langle \mathbf{H}\,,\,\mathbf{H}\right\rangle _{H_{\mathrm{p}}}=\int
\mathrm{d}^{3}\vec{r}\;\mathbf{H}(\vec{r})\cdot {\mu }(\vec{r})\cdot \mathbf{%
H}(\vec{r}),\qquad \left\langle \mathbf{E}\,,\,\mathbf{E}\right\rangle _{H_{%
\mathrm{q}}}=\int \mathrm{d}^{3}\vec{r}\;\mathbf{E}(\vec{r})\cdot {\epsilon }%
(\vec{r})\cdot \mathbf{E}(\vec{r}).
\end{equation}%
As a result
\begin{equation}
K_{\mathrm{p}}^{\dagger }=(2\sqrt{\pi })\nabla \times ,\quad K_{\mathrm{q}%
}^{\dagger }=\frac{1}{2\sqrt{\pi }}P_{\mathrm{curl}},\quad K={\mu }^{-1}(%
\vec{r})\cdot \nabla \times ,\quad K^{\dagger }={\epsilon }^{-1}(\vec{r}%
)\cdot \nabla \times ,  \label{ma2}
\end{equation}%
with $P_{\mathrm{curl}}$ the orthogonal projection of $(L^{2})^{3}$ onto $H_{%
\mathrm{curl}}$.

An alternative formulation of the general system (\ref{hpq1}) is suggested by
the example of the EM field, namely to consider (\ref{hpq3}) as generalized
material relations together with evolution equations
\begin{equation}
\partial _{t}%
\begin{pmatrix}
p \\
q%
\end{pmatrix}%
=%
\begin{pmatrix}
0 & -K_{\mathrm{q}}^{\mathrm{\dagger }} \\
K_{\mathrm{p}}^{\mathrm{\dagger }} & 0%
\end{pmatrix}%
\begin{pmatrix}
f_{\mathrm{p}} \\
f_{\mathrm{q}}%
\end{pmatrix}%
.  \label{hpq4a}
\end{equation}%
In turn, this suggests a natural modification incorporating dispersion and
dissipation by replacing (\ref{hpq3}) with:
\begin{equation}
f_{\mathrm{w}}(t)+\int_{0}^{\infty }\mathrm{d}\tau \,\chi _{\mathrm{w}}(\tau
)f_{\mathrm{w}}(t-\tau )=K_{\mathrm{w}}w(t),\text{ for }w=p,q,\ \mathrm{w}=%
\mathrm{p},\mathrm{q}.  \label{hpq5}
\end{equation}%
The TDD character of (\ref{hpq5}) comes from the \emph{operator valued
generalized susceptibilities }$\chi _{\mathrm{w}}$, $\mathrm{w}=\mathrm{p},%
\mathrm{q}$, the integrals of which explicitly satisfy the\emph{\ causality
condition: }values of $K_{\mathrm{w}}w\left( t\right) $ depend only on $f_{\mathrm{w}%
}\left( t^{\prime }\right) $ for times $t^{\prime }\leq t$.

Our main result is the following: \emph{Assume the susceptibilities }$\chi _{%
\mathrm{w}}$\emph{\ satisfy the following power dissipation condition (PDC),}
\begin{equation}
\func{Im}\left\{ \zeta \hat{\chi}_{\mathrm{w}}\left( \zeta \right) \right\} =%
\frac{1}{2\mathrm{i}}\left\{ \zeta \hat{\chi}_{\mathrm{w}}\left( \zeta
\right) -\zeta ^{\ast }\hat{\chi}_{\mathrm{w}}\left( \zeta \right) ^{\dagger
}\right\} \geq 0,\ \mathrm{w}=\mathrm{p},\mathrm{q},\text{ for all }\zeta
=\omega +i\eta ,\ \eta \geq 0,  \label{PDC}
\end{equation}%
\emph{for all $\zeta =\omega +i\eta $, $\eta \geq 0$, where $\hat{\chi}_{%
\mathrm{w}}$ is the Fourier-Laplace transform of $\chi _{\mathrm{w}}$:}
\begin{equation}
\hat{\chi}_{\mathrm{w}}\left( \zeta \right) =\frac{1}{\sqrt{2\mathrm{\pi }}}%
\int_{0}^{\infty }dt\,e^{\mathrm{i}\zeta t}\chi _{\mathrm{w}}\left( t\right)
,\emph{\ }\mathrm{w}=\mathrm{p},\mathrm{q}.
\end{equation}%
\emph{Then it is possible to construct a Hamiltonian extension to (\ref{hpq1}%
), which reduces to (\ref{hpq1}) in the limit of zero susceptibility, such that
the subsystem }$p,q$ \emph{evolves according to (\ref{hpq4a}, \ref%
{hpq5}).}

The extended Hamiltonian $\mathcal{A}\left( P,Q\right) $ is a function of
extended momentum $P$ and coordinate $Q$ variables, each taking values in a
Hilbert space $\mathcal{H}\supset H$, and has the same structure as (\ref%
{hpq1}), i.e.
\begin{equation}
\mathcal{A}\left( P,Q\right) =\frac{1}{2}\left\langle \mathcal{K}_{\mathrm{p}%
}P\,,\,\mathcal{K}_{\mathrm{p}}P\right\rangle +\frac{1}{2}\left\langle
\mathcal{K}_{\mathrm{q}}Q\,,\,\mathcal{K}_{\mathrm{q}}Q\right\rangle ,
\label{he6}
\end{equation}%
with $\mathcal{K}_{\mathrm{p}}$, $\mathcal{K}_{\mathrm{q}}$ closed operators
from $\mathcal{H}$ to $\mathcal{H}_{\mathrm{p}}\supset H_{\mathrm{p}}$, $%
\mathcal{H}_{\mathrm{q}}\supset H_{\mathrm{q}}$, which extend $K_{\mathrm{p}%
} $ and $K_{\mathrm{q}}$ respectively (see (\ref{he10}) below).

Before presenting the general construction, let us illustrate it with the
example of a linear TDD dielectric medium, described by the macroscopic
Maxwell equations without external charges and currents%
\begin{equation}
\partial _{t}\mathbf{D}=\nabla \times \mathbf{H},\qquad \partial _{t}\mathbf{%
B}=-\nabla \times \mathbf{E},\qquad \nabla \cdot \mathbf{B}=0,\qquad \nabla
\cdot \mathbf{D}=0,  \label{mxx1a}
\end{equation}%
in units with $c,\varepsilon _{0},\mu _{0}=1$. Here
\begin{equation}
\mathbf{D}=\mathbf{E}+4\mathrm{\pi }\mathbf{P},\qquad \mathbf{B}=\mathbf{H}+4%
\mathrm{\pi }\mathbf{M},  \label{mxx1b}
\end{equation}%
with the polarization $\mathbf{P}$ and magnetization $\mathbf{M}$ given by
linear response
\begin{equation}  \label{mxx3}
\mathbf{P}\left(\vec{r},t\right) =\int_{0}^{\infty }\mathrm{d}\tau \,\chi _{%
\mathrm{E}}\left( \vec{r},\tau \right) \mathbf{E}\left( \vec{r},t-\tau
\right) , \quad \mathbf{M}\left( \vec{r},t\right) =\int_{0}^{\infty }\mathrm{%
d}\tau \,\chi _{\mathrm{H}}\left( \vec{r},\tau \right) \mathbf{H}\left( \vec{%
r},t-\tau \right) .
\end{equation}%
The electric and magnetic susceptibilities should satisfy the PDC (\ref{PDC}%
) for each $\vec{r}$, and for simplicity we take them to be real valued
scalars. (The frequency domain susceptibilities $\widehat{\chi }_{\mathrm{F}%
}(\omega )$ may nonetheless be complex.)

Motivated by \cite{FS} and the Lamb model (see Fig.\ \ref{Lamb} below), we introduce canonical variables
\begin{equation}
P=\left( (4\pi )^{-1}\mathbf{A}(\vec{r}),\ \boldsymbol{\theta }_{\mathrm{E}}(%
\vec{r},s),\ \boldsymbol{\varphi }_{\mathrm{H}}(\vec{r},s)\right) ,\qquad
Q=\left( \mathbf{D}(\vec{r}),\ \boldsymbol{\varphi }_{\mathrm{E}}(\vec{r}%
,s),\ \boldsymbol{\theta }_{\mathrm{H}}(\vec{r},s)\right) ,
\end{equation}%
with $\mathbf{A},\mathbf{D}\in H_{\mathrm{curl}}$ and auxiliary vector
fields $\boldsymbol{\varphi }_{\mathrm{F}},\boldsymbol{\theta }_{\mathrm{F}}$%
, $\mathrm{F}=\mathrm{E},\mathrm{H}$, which are functions of $\vec{r}$ and
an auxiliary coordinate $-\infty <s<\infty $. For these variables we define
a Hamiltonian
\begin{equation}
\mathcal{A}\left( P,Q\right) =\mathcal{T}\left( P\right) +\mathcal{U}\left(
Q\right) ,  \label{fB5}
\end{equation}%
with,%
\begin{align}
\mathcal{T}\left( P\right) & =\frac{1}{2}\int \mathrm{d}^{3}\vec{r}
\left\vert 2 \sqrt{\pi} \, \nabla \times \frac{\mathbf{A}(\vec{r})}{4\pi }%
-\int_{-\infty }^{\infty }\mathrm{d}s\,\varsigma _{\mathrm{H}}\left( \vec{r}%
,s\right) \boldsymbol{\varphi }_{\mathrm{H}}(\vec{r},s)\right\vert ^{2}
\label{fB5a} \\
& \qquad \qquad  +\frac{1}{2}\int \mathrm{d}^{3}\vec{r}\int_{-\infty }^{\infty }%
\mathrm{d}s\left[ \left\vert \boldsymbol{\theta }_{\mathrm{E}}(\vec{r}%
,s)\right\vert ^{2}+\left\vert \partial _{s}\boldsymbol{\varphi }_{\mathrm{H}%
}(\vec{r},s)\right\vert ^{2}\right] ,  \notag
\end{align}%
\begin{align}
\mathcal{U}\left( Q\right) & =\frac{1}{2}\int \mathrm{d}^{3}\vec{r}
\left\vert \frac{1}{ 2 \sqrt{\pi}}\, \mathbf{D}(\vec{r})- \int_{-\infty
}^{\infty }\mathrm{d}s\,\varsigma _{\mathrm{E}}(\vec{r},s)\boldsymbol{%
\varphi }_{\mathrm{E}}(\vec{r},s)\right\vert ^{2}  \label{fB5b} \\
& \qquad \qquad+\frac{1}{2}\int \mathrm{d}^{3}\vec{r}\int_{-\infty }^{\infty }%
\mathrm{d}s\left[ \left\vert \partial _{s}\boldsymbol{\varphi }_{\mathrm{E}}(%
\vec{r},s)\right\vert ^{2}+\left\vert \boldsymbol{\theta }_{\mathrm{H}}(\vec{%
r},s)\right\vert ^{2}\right] ,  \notag
\end{align}%
where $\varsigma _{\mathrm{F}}$, $\mathrm{F}=\mathrm{E}, \mathrm{H}$, are
scalar functions to be specified below.

The resulting Hamilton equations of motion for the extended Maxwell system are:
\begin{gather}
\partial_t \frac{\mathbf{A}(\mathbf{r},t)}{4 \pi} = - \frac{1}{4\pi} \left [
\mathbf{D}(\mathbf{r},t) - 2 \sqrt{\pi} \left\langle \varsigma _{\mathrm{E}}
, \mathbf{\varphi}_{\mathrm{E}}\right\rangle_s \left( \mathbf{r},t\right)
\right ] , \\
\partial _{t}\mathbf{\varphi}_{\mathrm{H}}\left( \mathbf{r},s \right) =-%
\mathbf{\theta}_{\mathrm{H}}\left( \mathbf{r},s\right) ,  \notag \\
\partial_{t}\mathbf{\theta}_{\mathrm{E}}\left( \mathbf{r},s \right) =\frac{1%
}{ 2 \sqrt{\pi}}\, \varsigma_{\mathrm{E}}\left( \mathbf{r},s \right) \left [
\mathbf{D}(\mathbf{r},t) - 2 \sqrt{\pi} \left\langle \varsigma_{\mathrm{E}} ,%
\mathbf{\varphi}_{\mathrm{E}}\right\rangle_s \left( \mathbf{r},t\right) \right
] +\partial _{s}^{2}\mathbf{\varphi}_{\mathrm{E}}\left( \mathbf{r},s
\right) ,  \notag \\
\partial_t \mathbf{D}(\mathbf{r},t) = \nabla \times \left [\nabla \times \mathbf{A}(%
\mathbf{r},t) - 2\sqrt{\pi} \left\langle \varsigma _{\mathrm{H}},\mathbf{%
\varphi}_{\mathrm{H}}\right\rangle_s \left( \mathbf{r},t\right) \right ] , \\
\partial_{t}\mathbf{\theta}_{\mathrm{H}}\left( \mathbf{r},s \right) = -
\frac{1}{ 2 \sqrt{\pi}} \, \varsigma _{\mathrm{H}}\left( \mathbf{r},s \right)
\left [\nabla \times \mathbf{A}(\mathbf{r},t) - 2\sqrt{\pi} \left\langle
\varsigma _{\mathrm{H}},\mathbf{\varphi}_{\mathrm{H}}\right\rangle_s \left(
\mathbf{r},t\right) \right ] - \partial_s^2 \mathbf{\varphi}_{\mathrm{H}%
}\left( \mathbf{r},s \right) .  \notag \\
\partial_t \mathbf{\varphi_{\mathrm{E}}}\left( \mathbf{r},s \right) =
\mathbf{\theta}_{\mathrm{E}}\left( \mathbf{r},s\right) ,  \notag
\end{gather}
where
\begin{equation}
\left\langle \bullet, \bullet \right\rangle _{s}=\int \, \bullet \bullet \,%
\mathrm{d}s.  \label{fBa10}
\end{equation}
We make the identifications:
\begin{eqnarray}
\mathbf{B}(\mathbf{r},t) &=& \nabla \times \mathbf{A}(\mathbf{r},t) \\
\mathbf{H}\left( \mathbf{r},t\right) &=&\mathbf{B}\left( \mathbf{r},t\right)
-2\sqrt{\mathrm{\pi }}\left\langle \varsigma _{\mathrm{H}},\mathbf{\varphi}_{%
\mathrm{H}}\right\rangle_s \left( \mathbf{r},t\right) \ = \ \mathbf{B}\left(
\mathbf{r},t\right) -4\mathrm{\pi }\mathbf{M}\left( \mathbf{r},t\right) ,
\label{fBa11} \\
\mathbf{E}\left( \mathbf{r},t\right) &=&\mathbf{D}\left( \mathbf{r},t\right)
-2\sqrt{\mathrm{\pi }}\left\langle \varsigma _{\mathrm{E}},\mathbf{\varphi}
_{\mathrm{E}}\right\rangle_s \left( \mathbf{r},t\right) \ = \ \mathbf{D}%
\left( \mathbf{r},t\right) -4\mathrm{\pi }\mathbf{P}\left( \mathbf{r}%
,t\right) ,  \notag
\end{eqnarray}
with
\begin{equation}
\left\langle \varsigma _{\mathrm{H}},\mathbf{\varphi}_{\mathrm{H}%
}\right\rangle_s \left( \mathbf{r},t\right) =2\sqrt{\mathrm{\pi }}\, \mathbf{%
M}\left( \mathbf{r},t\right) ,\quad \left\langle \varsigma _{\mathrm{E}},%
\mathbf{\varphi} _{\mathrm{E}}\right\rangle_s \left( \mathbf{r},t\right) =2%
\sqrt{\mathrm{\pi }}\, \mathbf{P}\left( \mathbf{r},t\right) ,  \label{fBa9}
\end{equation}
resulting in the following equivalent system of extended Maxwell equations
\begin{gather}
\partial_{t}\mathbf{H}\left( \mathbf{r},t\right) =-\nabla \times \mathbf{E}%
\left( \mathbf{r},t\right) + 2 \sqrt{\pi} \, \left\langle \varsigma _{%
\mathrm{H}},\mathbf{\theta}_{\mathrm{H}}\right\rangle_s \left( \mathbf{r}%
,t\right) ,  \label{fBa7} \\
\partial _{t}\mathbf{\varphi}_{\mathrm{H}}\left( \mathbf{r},s \right) =-%
\mathbf{\theta}_{\mathrm{H}}\left( \mathbf{r},s\right) , \qquad \partial _{t}%
\mathbf{\theta}_{\mathrm{E}}\left( \mathbf{r},s \right) =\frac{1}{ 2 \sqrt{%
\pi}} \, \varsigma _{\mathrm{E}}\left( \mathbf{r},s \right) \mathbf{E}\left(
\mathbf{r},t\right) +\partial _{s}^{2}\mathbf{\varphi}_{\mathrm{E}}\left(
\mathbf{r},s \right) ,  \notag \\
\partial _{t}\mathbf{E}\left( \mathbf{r},t\right) =\nabla \times \mathbf{H}%
\left( \mathbf{r},t\right) -2 \sqrt{\pi }\left\langle \varsigma _{\mathrm{E}%
},\mathbf{\theta}_{\mathrm{E}}\right\rangle_s \left( \mathbf{r},t\right) ,
\label{fBa8} \\
\partial _{t}\mathbf{\theta}_{\mathrm{H}}\left( \mathbf{r},s\right) =-\frac{1%
}{ 2 \sqrt{\pi}} \,\varsigma _{\mathrm{H}}\left( \mathbf{r},s\right) \mathbf{%
H}\left( \mathbf{r},t\right) -\partial _{s}^{2}\mathbf{\varphi}_{\mathrm{H}%
}\left( \mathbf{r},s\right) ,\qquad \partial _{t}\mathbf{\varphi}_{\mathrm{E}%
}\left( \mathbf{r},s \right) = \mathbf{\theta}_{\mathrm{E}}\left( \mathbf{r}%
,s \right) .  \notag
\end{gather}

Combining the first order Hamilton equations of motion for $\boldsymbol{%
\varphi }_{\mathrm{F}}$ and $\boldsymbol{\theta }_{\mathrm{F}}$ into a
single second order equation for $\boldsymbol{\varphi }_{\mathrm{F}}$, $%
\mathrm{F}=\mathrm{E}, \mathrm{H}$, we obtain a driven wave equation
\begin{equation}
\left\{ \partial _{t}^{2}-\partial _{s}^{2}\right\} \boldsymbol{\varphi }_{%
\mathrm{F}}\left( \vec{r},s,t\right) =\frac{1}{ 2 \sqrt{\pi}} \, \varsigma _{%
\mathrm{F}}\left( \vec{r},s\right) \mathbf{F}(\vec{r},t), \quad \mathrm{F} =
\mathrm{E}, \mathrm{H}.  \label{fB7}
\end{equation}%
Assuming $\boldsymbol{\varphi }_{\mathrm{F}}$ to be at rest ($\boldsymbol{\varphi }_{\mathrm{F}} = \partial_t \boldsymbol{\varphi }_{\mathrm{F}} =0$) in the
distant past, the solution to (\ref{fB7}) is given by
\begin{equation}
\boldsymbol{\varphi }_{\mathrm{F}}\left( \vec{r},s,t\right) =\frac{1}{4
\sqrt{\pi}}\int_{0}^{\infty }\mathrm{d}\tau \int_{s-\tau }^{s+\tau }\mathrm{d%
}\sigma \,\varsigma _{\mathrm{F}}\left( \vec{r},\sigma \right) \mathbf{F}(%
\vec{r},t-\tau ),\quad \mathrm{F}=\mathrm{E},\mathrm{H},  \label{fb10}
\end{equation}%
implying with (\ref{mxx3}) and (\ref{fBa9}) the following expression for the
susceptibilities
\begin{equation}
\chi _{\mathrm{F}}\left( \vec{r},t\right) =\frac{1}{8 \pi}\int_{-\infty
}^{\infty }\mathrm{d}s\int_{s-t}^{s+t}\mathrm{d}\sigma \,\varsigma _{\mathrm{%
F}}\left( \vec{r},s\right) \varsigma _{\mathrm{F}}\left( \vec{r},\sigma
\right) ,\quad \mathrm{F}=\mathrm{E},\mathrm{H}.  \label{fb11}
\end{equation}
The key fact is that, due to the power dissipation condition \eqref{PDC}, it
is possible to invert \eqref{fb11} and write $\varsigma_{\mathrm{F}}$ as a
function of $\chi_{\mathrm{F}}$. An explicit solution is
\begin{equation}
\varsigma _{\mathrm{F}}\left( \vec{r},s\right) = \frac{2}{\sqrt[4]{2 \pi}}%
\int_{-\infty }^{\infty }\mathrm{d}\omega \,\mathrm{e}^{-\mathrm{i}\omega s}
\sqrt{\omega \mathrm{Im}\hat{\chi}_{\mathrm{F}}\left( \vec{r},\omega +%
\mathrm{i}0\right) },\quad \mathrm{F}=\mathrm{E},\mathrm{H}.  \label{fB6}
\end{equation}%
Note that $\varsigma _{\mathrm{F}}$ is real and invariant under $s\mapsto -s$.

The above discussion implies the following result on the extended system: \emph{Let the Hamiltonian (\ref{fB5}, \ref%
{fB5a}, \ref{fB5b}) be given with }$\varsigma _{\mathrm{F}}$\emph{\ defined
by (\ref{fB6}) for }$\chi _{\mathrm{F}}$\emph{\ which obey (\ref{PDC}). Then
for any solution to the Hamilton equations of motion with }$\varphi _{%
\mathrm{F}},\theta _{\mathrm{F}}\rightarrow 0$\emph{\ as }$t\rightarrow
-\infty $\emph{, the variables }$\mathbf{D}(\vec{r},t)$\emph{\ and }$\mathbf{%
B}(\vec{r},t)=\nabla \times \mathbf{A}(\vec{r},t)$\emph{\ evolve according
to the macroscopic Maxwell equations (\ref{mxx1a}-\ref{mxx3}).}

Based on the constructed TDD Hamiltonian (\ref{fB5}), we obtain an
expression for the energy density of the EM field and the medium
\begin{equation}
\mathcal{E}\left( \vec{r},t\right) =\frac{1}{8\mathrm{\pi }}\left\{
\left\vert \mathbf{E}\right\vert ^{2}+\left\vert \mathbf{H}\right\vert
^{2}\right\} \left( \vec{r},t\right) +\frac{1}{2}\left\{ \left\Vert \partial
_{s}\boldsymbol{\varphi }_{\mathrm{H}}\right\Vert _{s}^{2}+\left\Vert
\boldsymbol{\theta }_{\mathrm{H}}\right\Vert _{s}^{2}+\left\Vert \partial
_{s}\boldsymbol{\varphi }_{\mathrm{E}}\right\Vert _{s}^{2}+\left\Vert
\boldsymbol{\theta }_{\mathrm{E}}\right\Vert _{s}^{2}\right\} \left( \vec{r}%
,t\right) ,  \label{HS6}
\end{equation}%
where
\begin{equation}
\left\Vert \bullet \right\Vert _{s}^{2}=\int \left\vert \bullet \right\vert
^{2}\,\mathrm{d}s.
\end{equation}
This results in the conservation law
\begin{equation}
\partial _{t}\mathcal{E}+\nabla \cdot \mathbf{S}=0,  \label{HS9}
\end{equation}%
with the familiar Poynting vector for the energy flux
\begin{equation}
\mathbf{S}\left( \vec{r},t\right) =\frac{1}{4\mathrm{\pi }}\mathbf{E}\left(
\vec{r},t\right) \times \mathbf{H}\left( \vec{r},t\right) .  \label{HS8}
\end{equation}%
These identities follow from (\ref{fB5}-\ref{fB5b}) and the general theory
of Hamiltonian fields \cite{LandauLif1}.

When the interaction $\varsigma $ is set to zero, the EM and auxiliary
fields decouple and (\ref{HS6}) reduces to%
\begin{equation}
\mathcal{E}_{0}\left( \vec{r},t\right) =\mathcal{E}_{\mathrm{EM}}\left( \vec{%
r},t\right) +\mathcal{E}_{\mathrm{S}}\left( \vec{r},t\right)  \label{HS9a}
\end{equation}%
with the energy density of the EM field
\begin{equation}
\mathcal{E}_{\mathrm{EM}}\left( \vec{r},t\right) =\frac{1}{8\mathrm{\pi }}%
\left\vert \mathbf{D}\right\vert ^{2}\left( \vec{r},t\right) +\frac{1}{8%
\mathrm{\pi }}\left\vert \nabla \times \mathbf{A}\right\vert ^{2}\left( \vec{%
r},t\right) ,
\end{equation}%
and the energy density of the auxiliary fields
\begin{equation}
\mathcal{E}_{\mathrm{S}}\left( \vec{r},t\right) =\frac{1}{2}\left\{
\left\Vert \partial _{s}\boldsymbol{\varphi }_{\mathrm{H}}\right\Vert
_{s}^{2}+\left\Vert \boldsymbol{\theta }_{\mathrm{H}}\right\Vert
_{s}+\left\Vert \partial _{s}\boldsymbol{\varphi }_{\mathrm{E}}\right\Vert
_{s}^{2}+\left\Vert \boldsymbol{\theta }_{\mathrm{E}}\right\Vert
_{s}^{2}\right\} \left( \vec{r},t\right) .  \label{HS12}
\end{equation}%
Subtracting (\ref{HS9a}) from (\ref{HS6}) gives the energy shift due to the
interaction of the EM field and the matter
\begin{equation}
\delta \mathcal{E}\left( \vec{r},t\right) =\frac{1}{8\mathrm{\pi }} \left \{
\left\vert \mathbf{E}\left( \mathbf{r},t\right) \right\vert ^{2}-\left\vert
\mathbf{D}\left( \mathbf{r},t\right) \right\vert ^{2}+\left\vert \mathbf{H}%
\left( \mathbf{r},t\right) \right\vert ^{2}-\left\vert \mathbf{B}\left(
\mathbf{r},t\right) \right\vert ^{2} \right \}.
\end{equation}

In general, it is not possible to give an expression for the energy density $%
\mathcal{E}_{\mathrm{S}}\left( \vec{r},t\right) $ of the medium in terms of
the instantaneous EM fields $\mathbf{E}(\vec{r},t)$ and $\mathbf{H}(\vec{r}%
,t)$. However, using (\ref{fb10}, \ref{fb11}) and the equations of motion we
have calculated that
\begin{align}
\partial_t \, \frac{1}{2} \left\{ \Vert \partial _{s}\boldsymbol{\varphi }_{%
\mathrm{F}}\Vert _{\mathrm{s}}^{2}+\Vert \boldsymbol{\theta }_{\mathrm{F}%
}\Vert _{\mathrm{s}}^{2}\right\} (\vec{r},t)&= \frac{1}{2 \sqrt{\pi}} \left
[ \partial_t \langle \varsigma_{\mathrm{F}} , \boldsymbol{\varphi}_{\mathrm{F%
}} \rangle (\vec{r}, t) \right ] \cdot \mathbf{F}(\vec{r},t)  \label{last} \\
&=\left[ \partial _{t}\mathbf{P}(\vec{r},t )\right] \cdot \mathbf{E}(\vec{r}%
,t ), \quad \mathrm{F} = \mathrm{E},  \label{lastE}
\end{align}
with a similar expression for $\mathrm{F}=\mathrm{H}$.

The result (\ref{lastE}) is the usual expression for the rate of change
the density of EM energy stored in a dielectric. For a wave packet $\mathbf{E%
}(\vec{r},t)=\mathrm{Re}\,\left\{ \mathrm{e}^{-\mathrm{i}\omega _{0}t}%
\mathbf{E}_{0}(\vec{r},t)\right\} $ with $\mathbf{E}_{0}(\vec{r},t)$ a slowly
varying function of $t$, we have
\begin{equation}
\partial _{t}\mathbf{P}(\vec{r},t)\ \approx \ \sqrt{2\pi }\mathrm{Re}\left\{
-\mathrm{i}\omega _{0}\widehat{\chi }_{\mathrm{E}}(\vec{r},\omega _{0})%
\mathrm{e}^{-\mathrm{i}\omega _{0}t}\mathbf{E}_{0}(\vec{r},t)+\frac{\mathrm{d%
}\omega _{0}\widehat{\chi }_{\mathrm{E}}(\vec{r},\omega _{0})}{\mathrm{d}%
\omega _{0}}\mathrm{e}^{-\mathrm{i}\omega _{0}t}\partial _{t}\mathbf{E}_{0}(%
\vec{r},t)\right\} \;,
\end{equation}%
and thus
\begin{multline}
\left[ \partial _{t}\mathbf{P}(\vec{r},t)\right] \cdot \mathbf{E}(\vec{r},t)
\\
\approx \ \frac{\sqrt{2\pi }}{2}\mathrm{Re}\,\left\{ -\mathrm{i}\omega _{0}%
\widehat{\chi }_{\mathrm{E}}(\vec{r},\omega _{0})|\mathbf{E}_{0}(\vec{r}%
,t)|^{2}+\frac{\mathrm{d}\omega _{0}\widehat{\chi }_{\mathrm{E}}(\vec{r}%
,\omega _{0})}{\mathrm{d}\omega _{0}}\left[ \partial _{t}\mathbf{E}_{0}(\vec{%
r},t)\right] \cdot \mathbf{E}_{0}^{\ast }(\vec{r},t)\right\}  \\
+\text{ terms with a factor of $\mathrm{e}^{-\mathrm{i}\omega _{0}t}$ or $%
\mathrm{e}^{\mathrm{i}\omega _{0}t}$.}
\end{multline}%
If we consider the \emph{time averaged} power density, averaged over a time
scale longer than $1/\omega _{0}$, the terms with oscillatory factors are
very small and we have (with $\overline{\ \bullet }$ denoting time
averaging):
\begin{multline}
\overline{\ \left[ \partial _{t}\mathbf{P}(\vec{r},t)\right] \cdot \mathbf{E}%
(\vec{r},t)}  \label{Brioullin} \\
\approx \sqrt{\frac{\pi }{2}}\left\{ \frac{1}{2}\left[ \frac{\mathrm{d}\ }{%
\mathrm{d}\omega _{0}}\omega _{0}\mathrm{Re}\widehat{\chi }_{\mathrm{E}}(%
\vec{r},\omega _{0})\right] \,\partial _{t}|\mathbf{E}_0(\vec{r}%
,t)|^{2}+\omega _{0}\mathrm{Im}\widehat{\chi }_{\mathrm{E}}(\vec{r},\omega
_{0})|\mathbf{E}_0(\vec{r},t)|^{2}\right\} \;,
\end{multline}%
where we have assumed for simplicity that the slowly varying function $%
\mathbf{E}_{0}(\vec{r},t)$ is real. The first term on the r.h.s. of \eqref{Brioullin} is a
total derivative, which when integrated gives the Brillouin formula for the
time averaged energy density in a material medium (see \cite[Section
80]{LandLifEM}):
\begin{equation}
\frac{\sqrt{2\pi }}{4}\left[ \frac{\mathrm{d}\ }{\mathrm{d}\omega _{0}}%
\omega _{0}\mathrm{Re}\widehat{\chi }_{\mathrm{E}}(\vec{r},\omega _{0})%
\right] \,|\mathbf{E}_0(\vec{r},t)|^{2}\;,
\end{equation}%
which neglects losses entirely, and hence is useful only if $\mathrm{Im}%
\widehat{\chi }_{\mathrm{E}}(\vec{r},\omega _{0})$ is zero or so small as to be
irrelevant. The second term of \eqref{Brioullin} is strictly positive and
gives the contribution
\begin{equation}
\sqrt{\frac{\pi }{2}}\omega _{0}\mathrm{Im}\widehat{\chi }_{\mathrm{E}}(\vec{%
r},\omega _{0})\int_{-\infty }^{t}|\mathbf{E}_0(\vec{r},\tau )|^{2}\mathrm{d}%
\tau \;
\end{equation}%
to the time averaged material energy density, which is non-decreasing in
time and incorporates losses in an approximate way.

For a general system of the form (\ref{hpq4a}, \ref{hpq5}, %
\ref{PDC}), a TDD Hamiltonian  can be constructed in the same way. We define canonical variables
\begin{equation}
P=\left( p,\ \theta _{\mathrm{q}},\ \varphi _{\mathrm{p}}\right) ,\qquad
Q=\left( q,\ \varphi _{\mathrm{q}},\ \theta _{\mathrm{p}}\right) ,
\end{equation}%
with $\theta _{\mathrm{w}}(s),\varphi _{\mathrm{w}}(s)$ functions of an
auxiliary coordinate $-\infty <s<\infty $, taking values in the Hilbert
spaces $H_{\mathrm{w}}$, $\mathrm{w}=\mathrm{p},\mathrm{q}$. If $\chi _{%
\mathrm{w}}(t)=\chi _{\mathrm{w}}(t)^{\dagger }$ for all $t\geq 0$, the
Hamiltonian is of the form (\ref{he6}), with
\begin{eqnarray}
\mathcal{K}_{\mathrm{p}}P &=&\left( K_{\mathrm{p}}p-\int_{-\infty }^{\infty }%
\mathrm{d}s\,\varsigma _{\mathrm{p}}(s)\varphi _{\mathrm{p}}(s),\ \theta _{%
\mathrm{q}},\ \partial _{s}\varphi _{\mathrm{p}}\right) ,  \label{he10} \\
\mathcal{K}_{\mathrm{q}}Q &=&\left( K_{\mathrm{q}}q-\int_{-\infty }^{\infty }%
\mathrm{d}s\,\varsigma _{\mathrm{q}}(s)\varphi _{\mathrm{q}}(s),\ \partial
_{s}\varphi _{\mathrm{q}},\ \theta _{\mathrm{p}}\right) ,  \notag
\end{eqnarray}%
where%
\begin{equation}
\hat{\varsigma}_{\mathrm{w}}(\omega )=(2\mathrm{\pi })^{-1/4}\sqrt{2\omega
\mathrm{Im}\chi _{\mathrm{w}}(\omega +\mathrm{i}0)}, \quad \mathrm{w}=\mathrm{p},%
\mathrm{q},
\end{equation}
with $\sqrt{\bullet}$ the operator square root. 

In particular, we can handle it this way:
\textit{i.}) Non-isotropic media, provided the tensors $\chi _{\mathrm{w}}$
are real symmetric. (Gyrotropy could in principle be handled by a more
involved construction with terms mixing momenta and coordinates.) \textit{ii.%
}) Space dispersion, in which case terms depending on $\nabla \boldsymbol{%
\varphi }$ and $\nabla \boldsymbol{\theta }$ appear in the Hamiltonian.
Details of the abstract construction and further examples will be given in
forthcoming work \cite{FS2}.

\section{Discussion and comparison with prior work}

The need for a Hamiltonian description of a dissipative system has long been
known, having been emphasized by Morse and Feshbach \cite[Ch. 3.2.]{Importance}
forty years ago. They constructed, for a damped oscillator, an artificial
Hamiltonian based on a ``mirror-image'' trick, incorporating a second
oscillator with negative friction. The resulting Hamiltonian is quite
un-physical: it is unbounded from below and under time reversal the oscillator
is transformed into its ``mirror-image.'' The artificial nature of this
construction was described in \cite[Ch. 3.2.]{Importance}: ``By this arbitrary
trick we are able to handle dissipative systems as though they were
conservative. This is not very satisfactory if an alternate method of solution
is known...''

The Hamiltonian we construct for TDD media can be viewed as a quite general
``satisfactory solution'' to the problem posed in \cite[Ch. 3.2.]{Importance}
since \emph{we do not introduce negative friction} and, in particular, we do
not make use of ``mirror-images.'' Instead we couple a given TDD system to
an effective model for the normal modes of the underlying medium. For the
combined system we give a non-negative Hamiltonian with a transparent
interpretation as the system energy. As regards the underlying microscopic
theory, this is an effective Hamiltonian for those modes well approximated
by linear response.

\begin{figure}
[ptb]
\begin{center}
\includegraphics[
height=2.4722in, width=3.2877in ] {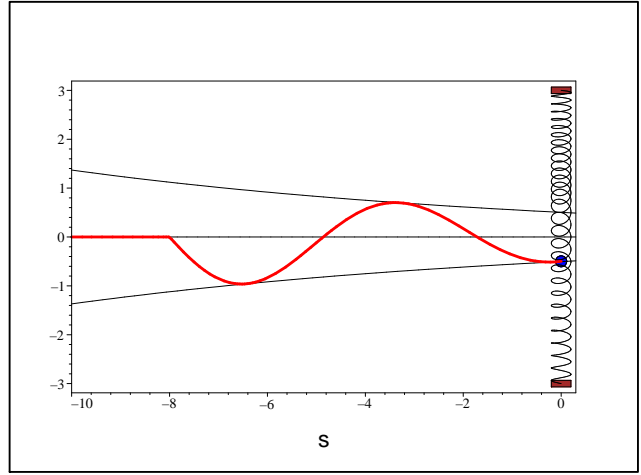} \caption{The Lamb model,
introduced in \cite{Lamb} to describe radiation damping, is a point mass
attached to an infinite elastic string and a Hook's
law spring. The point mass evolves as a classical linearly damped oscillator.}%
\label{Lamb}
\end{center}
\end{figure}

The evolution equations of the proposed theory come from a Hamiltonian and
are thus time reversible. Nonetheless, an irreversible motion of the TDD
system stems from the infinite heat capacity of the auxiliary system. This
is demonstrated in its simplest form by the damped harmonic oscillator 
\begin{equation}
m \partial_t^2q(t) + \gamma \partial_t q(t) + k q(t) \ = \ 0 ,
\end{equation}
which results from the TDD system
\begin{equation}
\partial_t \begin{pmatrix}
p \\
q
           \end{pmatrix} \ = \  \begin{pmatrix} - f_{\mathrm{q}} \\ 
	   f_{\mathrm{p}}
                                \end{pmatrix} \ , \quad \begin{pmatrix}
				f_{\mathrm{p}}(t) + \gamma \int_0^\infty  f_{\mathrm{p}}(t - \tau) \mathrm{d} \tau \\
				f_{\mathrm{q}}(t) 
				\end{pmatrix}  \ = \  \begin{pmatrix} \frac{1}{\sqrt{m}} p(t) \\
				\sqrt{k} q(t) \end{pmatrix} ,
\end{equation} with $m,k,\gamma >0$. The construction presented above reproduces in this simple case a model due to Lamb in 1900 \cite{Lamb} ---\ see
Fig.\ \ref{Lamb} ---\ in which the energy of an oscillator escapes to
infinity along an attached flexible string. The theory proposed here illustrates that, from the standpoint of thermodynamics, dissipation in classical linear response is an idealization
which assumes infinite heat capacity of (hidden) degrees of freedom.

In general, the auxiliary system, described by the
fields $\left\{ \varphi_{\mathrm{p}}, \varphi_{\mathrm{q}} ,
\theta_{\mathrm{p}}, \theta_{\mathrm{q}} \right\} $, is governed by a Hamiltonian $%
\mathrm{H}_{\mathrm{hb}}(\varphi ,\theta )$ of the following simple and
universal form%
\begin{equation}
\mathrm{H}_{\mathrm{hb}}(\varphi ,\theta )\ =\ \frac{1}{2}\int_{0}^{\infty }%
\left[ \left\Vert \theta \left( s\right) \right\Vert _{H_{\mathrm{p}} \oplus H_{\mathrm{q}} }^{2}+\left\Vert
\partial _{s}\varphi \left( s\right) \right\Vert _{H_{\mathrm{p}} \oplus H_{\mathrm{q}}}^{2}\right] \,\mathrm{d}%
s.  \label{Hhb1}
\end{equation}%
The Hamiltonian $\mathrm{H}_{\mathrm{hb}}(\varphi ,\theta )\ $in (\ref{Hhb1}%
) is a \emph{canonical heat bath} (justifying the
index $\mathrm{hb}$) as described in \cite[Section 2]{JP2}, \cite[Section 2]{RB}.

The physical concept of an ideal or canonical heat bath originates in 
thermodynamics. Statistical mechanical models at the
mathematical level of rigor were introduced, motivated, and described rather
recently (to our best knowledge), see \cite[Section 1]{KKS}, \cite[Section 2]%
{JP2}, \cite[Section 2]{RB} and references therein. The consensus among the
references, on the basis of statistical mechanics, is that the generator of motion
for a canonical heat bath must be
$\mathrm{i} \times $ a self-adjoint operator with absolutely continuous spectrum with
no gaps, i.e. the spectrum must be the entire real line $\mathbb{R}$, and
the spectrum must be of a uniform multiplicity. These requirements lead to a
system unitarily equivalent to the universal form Hamiltonian (\ref{Hhb1}), \cite[%
Section 2]{JP2}.

General statistical mechanics considerations indicate that for a system to
behave according the thermodynamics it must be properly coupled to the heat
bath. In particular, the coupling should involve all modes of the heat
bath, \cite[Section 1]{KKS}, \cite[Section 2]{JP2}, \cite[Section 2]{RB}. In
our Hamiltonian setting the coupling is essentially 
$\langle  K_{\mathrm{w}}w, T_{\mathrm{w}} \varphi_{\mathrm{w}} \rangle $, which is the \emph{dipole approximation}, \cite[Section
1,2]{RB}, and the condition of 
coupling to all modes is the following constraint on the dissipation:
\begin{equation}
\overline{
\bigcup_{\omega \in \mathbb{R}}\limfunc{Ran}\mathrm{Im}\widehat{\chi }_{\mathrm{w}}(\omega )}%
=H_{\mathrm{w}}, \quad \mathrm{w}= \mathrm{p,q}.
\end{equation}

There is some relation in spirit between our theory and a recently proposed
hydrodynamic theory (HT) \cite{Liu1}, \cite{Liu2}, \cite{Liu3}. Both
theories are self contained, macroscopic, and make no assumption on the
underlying microstructure. However the theory proposed here, unlike the HT,
makes no use of parameters other than the susceptibilities of linear
response theory. Furthermore, the present theory is truly conservative, with dissipative
effects modeled by effectively irreversible energy transport to an auxiliary
system, which may be conceived of as constructed from flexible strings. In
contrast, the HT makes use of explicitly dissipative, nonconservative
equations similar to those of Navier-Stokes.

A deeper relation can be found between the approach described here and 
the well known dilation theory which, in certain cases, provides a
treatment for dissipation and resonance phenomena. We first give a brief
account of the dilation theory, based on B. Pavlov's extensive review \cite%
{Pav1} as well as his more recent work \cite{Pav2}. For more detailed
exposition on the subject we refer the reader to \cite{Pav1}, \cite{Pav2} and
references therein.

The dilation theory was the first rather general approach to the
construction of a spectral theory for dissipative operators. It is based on
an abstract version of the Lax-Phillips scattering theory \cite{LaxPh} which
assumes that there are \textit{i.}) a dynamical unitary evolution group $U_{t}=\mathrm{%
e}^{\mathrm{i}\Omega t}$ in a Hilbert space $H$ where $\Omega $ is a
self-adjoint operator in $H$; \textit{ii.}) an \textquotedblleft
\emph{incoming}"
subspace $D_{-}\subset H$ invariant with respect to the semi-group $U_{t}$, $%
t<0$, and an \textquotedblleft \emph{outgoing}" subspace $D_{+\text{ }%
}\subset H$ invariant with respect to the semi-group $U_{t}$, $t>0$. The
invariant subspaces (also called \emph{scattering channels}) $D_{\pm \text{ }%
}$ \emph{are assumed to be orthogonal}. Then one introduces the
\textquotedblleft \emph{observation}" subspace $K=H\ominus \left( D_{-}\oplus
D_{+}\right) $ which is coinvariant in the sense that the restriction of
$U_{t}$, $t>0$, to $K$ is a well defined semigroup on its own, namely for $t>0$
\begin{equation}
Z_{t}=\left. P_{K}U_{t}\right\vert _{K}=\mathrm{e}^{\mathrm{i}Bt},\text{ where
}P_{K}\,\text{is the orthogonal projection on }K\text{.}  \label{zp1}
\end{equation}

In many interesting cases the generator $B$ of the
semigroup $Z_{t}$ is dissipative, i.e. $\func{Im}B\geq 0$ or, even, $\func{Im%
}B>0$. So the relation (\ref{zp1}) provides an interesting scenario within
the Lax-Phillips scattering theory for the rise of a dissipative operator.
The dilation theory yields the spectral theory through the construction of
generalized eigenmodes (the scattering theory) provided, of course, the
conditions discussed above are satisfied.

Looking at the dilation theory from the point of view of open (dispersive
and dissipative) systems, one can ask if the theory allows to find the
unitary group $U_{t}=\mathrm{e}^{\mathrm{i}\Omega t}$ or, equivalently, the
self-adjoint operator $\Omega $ being given the dissipative operator $B$?
The answer is positive for rather large class of dissipative operators $B$.
For example, if $B=\Omega _{0}+\mathrm{i}a$, where $\Omega _{0}$ is self-adjoint
and $a\geq 0$ is bounded, a unique minimal dilation and its eigenmodes
can effectively constructed, \cite[Theorem 3]{Pav2}.

Hence when the dilation theory applies it provides a solid foundation for
spectral studies. Unfortunately, the dilation theory does not apply to many
important physical problems simply because its initial assumptions on the
nature of the dissipation are too restrictive. For systems described by
evolution equations (\ref{hpq4a})-(\ref{hpq5}) the dissipation always comes
with the dispersion, and the dilation theory does not apply. Indeed, the
most general form for a linear causal time-homogeneous \emph{open system}, as analyzed in \cite{FS}, is
\begin{equation}
m\partial _{t}v\left( t\right) =-\mathrm{i}Av\left( t\right)
-\int_{0}^{\infty }a\left( \tau \right) v\left( t-\tau \right) \,d\tau
+f\left( t\right) ,\ v\left( t\right) \in H_{0}  \label{os1}
\end{equation}%
where $H_{0}$ is a Hilbert space, $m>0$ and $\Omega $ are self-adjoint
operators in $H_{0}$, $f\left( t\right) $ is an external force and $a\left(
t\right) $ is the \emph{friction function}, subject to a power dissipation
condition%
\begin{equation}
\func{Re}\int_{0}^{\infty }\int_{0}^{\infty }\overline{v\left( t\right) }%
a\left( \tau \right) v\left( t-\tau \right) \,dtd\tau \geq 0.  \label{os2}
\end{equation}%
\emph{Only in the very special case of (\ref{os1}) when the friction is
instantaneous (Markovian), i.e. }$a\left( t\right) =a_{0}\delta \left(
t\right) $\emph{, can one use the dilation theory as in \cite[Theorem 3]%
{Pav2}}. For many well studied dielectric media, such as Lorentz or Debye, not to mention
media with generic frequency dependent electric susceptibilities, the
relevant friction functions are not instantaneous. For such systems one
must use a more general approach, such as developed in \cite{FS} and
extended here and in \cite{FS2}.

It is interesting to point out, however, that coupling to a canonical heat bath
(\ref{Hhb1}) may be interpreted, as in the Lamb model, as ``attaching an
elastic string" at any point of loss. The attached strings are analogous to
the scattering channels of the dilation theory, with the only difference
being that our ``strings" are coupled in more general ways than in the
dilation theory. Thus, one can view our TDD Hamiltonian as a natural
generalization of the constructions of the dilation theory.



\textbf{Acknowledgment}: \emph{We thank Lars Jonsson and Ilya Vitebskiy for
useful discussions. Support under AFOSR grant FA9550-04-1-0359 is gratefully
acknowledged.}

\end{document}